# Tracing magnetic atom diffusion with annealing at the interface between CoMn alloy and MnGa layer by X-ray magnetic circular dichroism

Jun Okabayashi,[1*] Kazuya Z. Suzuki,[2,3] and Shigemi Mizukami[3,4]

[1]*Research Center for Spectrochemistry, The University of Tokyo, Bunkyo-ku, Tokyo 113-0033, Japan*
[2]*Advanced Science Research Center, Japan Atomic Energy Agency, Tokai, Ibaraki 319-1195, Japan*
[3]*WPI-Advanced Institute for Materials Research, Tohoku University, Sendai 980-8579, Japan*
[4]*Center for Science and Innovation in Spintronics (CSIS), Core Research Cluster (CRC), Tohoku University, Sendai 980-8577, Japan*
*e-mail address: jun@chem.s.u-tokyo.ac.jp

(2022.10.28)

The magnetic atom diffusion at the interface between CoMn alloy and MnGa layer with annealing is studied using x-ray magnetic circular dichroism (XMCD) analysis. We found that the spins in bcc CoMn are coupled parallel to those in perpendicularly magnetized MnGa layer under the as-grown conditions, while the post-annealing modulates the interfacial magnetic coupling to antiferromagnetic in Co. The element-specific hysteresis curves at each absorption edge revealed the large coercive fields in Mn and Co through the exchange coupling with MnGa. After the annealing process, the changes of XMCD spectral line shapes are related to the interfacial reactions promoting the formation of Mn and $Co_2MnGa$ layers, which is deduced from the analysis of transmission electron microscopy. The interfacial diffusion of Mn atoms modulates the magnetic exchange coupling between Mn and Co sites and reverses the direction of perpendicular magnetization.

Controlling the interfacial properties is an essential issue for developing the spintronics materials because the symmetry-broken interfaces in the magnetic thin films provide some unique properties such as perpendicular magnetic anisotropy (PMA) and exchange coupling between different layers. It directly links to simultaneously achieve high storage density and low power consumption for operation in nano-scaled spintronic devices [1-3]. Large PMA is also required for the high-speed perpendicular switching in tunnel magnetoresistance (TMR) devices. For the developments of the magnetic tunnel junctions in the TMR devices, the investigations of interfacial abruptness and chemical compositions are strongly required because the TMR ratio is directly influenced by the qualities of interfacial atomic structures.

As promising materials for TMR devices, tetragonal $Mn_{3-\delta}Ga$ alloys, where $\delta$ is 0, 1, and 2,

have been extensively investigated because this alloy exhibits high PMA energies of over 1.0 MJ/m$^3$, low magnetic damping constants, and low saturation magnetizations originating from their ferrimagnetic properties that cause antiferromagnetic coupling at the different Mn sites [4-11]. Using the advantage of Mn$_{3-\delta}$Ga as a hard magnetic film, the deposition of other ferromagnetic materials on Mn$_{3-\delta}$Ga layers can be used to induce perpendicular magnetization through exchange interactions [12,13]. Ultra-thin Fe or Co layers deposited on Mn$_{3-\delta}$Ga couples ferromagnetically or antiferromagnetically, respectively, after the annealing [14]. Further, the annealing proceeds the Mn atomic diffusion from the interface and the appearance of secondary phase is inevitable. The abruptness and element-specific magnetic properties at the interfaces between Mn$_{3-\delta}$Ga and transition-metal (TM) alloy layers must be clarified explicitly. In particular, TMR ratios after the annealing found to appear higher than those in as-grown samples. There is a trade-off relation between high temperature annealing for high crystallinity and the interfacial reaction. Alternative approaches for positive use of secondary formation might bring the interfacial novel properties as future material science.

Other candidates for the magnetic switching layer combined with Mn$_{3-\delta}$Ga are the metastable body-centered-cubic (bcc) (Co,Mn)-based alloy which exhibits high TMR ratio and is compatible to the MgO insulating barrier layer [15-18]. In particular, perpendicular magnetization switching using Mn-based alloys is strongly demanded for high TMR spintronic devices [19]. However, the interfacial diffusion of Mn atoms has to be pursued explicitly. Compared with the case of Co/MnGa interface, the amounts of Mn atoms increase which modulates the local environment around the Co sites. The ferro- or antiferromagnetic exchange coupling in (Co,Mn)-based alloys has been extensively investigated depending on the compositions and crystalline structures [20-27]. Therefore, the element-specific characterizations through the annealing process are necessary to understand the ability of interfacial atomic controlling, which results in the development of the Mn-based spintronics materials [4].

For above motivations, x-ray magnetic circular dichroism (XMCD) can be a powerful tool for investigating the element-specific spin and orbital magnetic moments, and it enables measurement of element-specific magnetic-field dependence in XMCD hysteresis (*M-H*) curves at the fixed photon energy [28]. In a previous report, an XMCD study that was used to examine an (Fe,Co)/MnGa interface was presented, and the ferromagnetic coupling in the as-grown conditions was discussed [12]. The Co/MnGa interfaces with and without annealing processes are also examined by XMCD hysteresis curves. Antiparallel coupling between Co and MnGa can be directly detected by XMCD.

In this paper, we present the x-ray absorption spectroscopy (XAS), XMCD, and transmission electron microscopy (TEM) with energy dispersive x-ray (EDX) analysis to investigate the element-specific magnetic and structural properties at the interfaces between bcc

CoMn alloy and MnGa layer with annealing process. In particular, we discuss the interfacial diffusion and magnetic coupling, which may be ferromagnetic or antiferromagnetic depending on the annealing of the samples.

The samples were prepared by magnetron sputtering on MgO (001) substrates. After the annealing at 700°C, the 40-nm-thick Cr buffer layer was deposited at room temperature (RT), then annealed at 700°C. On this surface, 30-nm-thick $Mn_{61}Ga_{39}$ (MnGa) alloys was grown at RT with the post-annealing at 500°C. Then, 1-nm-thick $Co_{80}Mn_{20}$ (CoMn) layer was deposited with 2-nm-thick MgO capping layer. The sputtering growth was performed under the base pressure of $10^{-7}$ Pa and the Ar gas pressure of 0.1 Pa. Three kinds of samples of as-grown and post-annealed at 250 and 350°C were prepared. The depth-profile analysis for the sample after the annealing at 350°C was conducted by cross-sectional TEM and EDX analysis using the $K$-edge fluorescence in each element.

The XMCD analysis was performed at BL-7A in the Photon Factory at the High-Energy Accelerator Research Organization (KEK-PF). The photon helicity was fixed, and a magnetic field of ±1.2 T was applied parallel to the incident polarized soft X-ray beam and defined as $\mu^+$ and $\mu^-$ spectra. The total electron yield (TEY) mode was adopted, and all measurements were performed at RT. The XAS and XMCD measurement geometries were set to normal incidence, so that both the photon helicity and the magnetic field were normal to the surface, enabling measurements of the absorption processes involving the perpendicularly magnetized states. The element-specific $M$-$H$ curves were collected by scanning the magnetic field at each $L_3$-edge photon energy.

Figure 1 shows the wide-range XAS in $(\mu^++\mu^-)/2$ including Mn and Co $L$-edges for CoMn/MnGa with annealing temperature dependence and the reference spectrum of bcc CoMn film. The spectra were normalized at the Co $L_3$-edge peak intensity to be 1 in order to deduce the intensity ratio between Mn and Co. For the 30-nm-thick CoMn film without MnGa as a reference [18], the intensity ratio of Mn to Co is estimated to be 0.55 from the spectral intensity in Fig. 1. Considering the difference in photo-ionization cross section, the peaks of Mn $L$-edge are pronounced with a factor of 1.63 compared with the case of Co. For the CoMn/MnGa, the spectral intensity ratios of Mn to Co are estimated to be 1.68, 2.08, and 3.22 for as-grown, 250°C, and 350°C annealing, respectively, from Fig. 1. At the as-grown case, since the Mn $L$-edge peaks include the components of both CoMn and MnGa, the contribution from MnGa can be estimated by subtracting that from CoMn to be 1.13 (=1.68-0.55). After the annealing, the Mn intensity is drastically enhanced because of the segregation of Mn atoms onto the surface region. At the 350°C annealing case, the composition ratio of Mn to Co becomes 1.98 (=3.22/1.63), which is discussed later with the results of TEM and XMCD as the formation of Mn and $Co_2MnGa$ layers.

In order to trace the depth profile by annealing, the TEM and element-specific EDX images are shown in Fig. 2. By the annealing at 350°C, the bcc structures are maintained for all layers from the lattice image. High-Angle Annular Dark-Field Scanning TEM (HAADF-STEM) images for each element are also shown in Fig. 2(a). The interfacial mixing occurs as shown in the Mn, Co, and Ga profiles. The Mn atom diffusion can be detected onto the MgO interface and the peak appears at the 2-nm distance. The 1-nm-thick CoMn alloy layer is modulated to the composition of $Co_2MnGa$ beneath 2.5-3.5 nm from the surface. The thickness of 1 nm can be deduced with the diffusive interface from EDX depth profile in Fig. 2(b). The HAADF-STEM images of Co and Ga also display the interfacial diffusion. These results are consistent with the intensity profile of XAS in Fig. 1; the enhancement of Mn intensity in XAS comes from the segregation of Mn layer beneath 1.5-2.5 nm from the surface. The spike-like feature in the depth profile of Mn and Ga in the bottom MnGa layer originates from the perfect layer-by-layer ordering in Mn and Ga atoms.

The XAS and XMCD of Mn and Co $L_{2,3}$-edges for as-grown, 250 and 350°C annealing are shown in Fig. 3. The XAS intensities in each absorption edge are normalized as the post-edge to be one to pronounce the XAS and XMCD line shapes at each absorption edge. Metallic line shapes are clearly observed. With increasing the annealing temperature, the XAS line shapes in Mn are almost identical. On the other hand, in Co, a small satellite structure appears in higher energy side of 782 eV, which is also detected in the case of Co/MnGa interface [12] and suggested the chemical bond with an anion element in the Co-based Heusler alloys [28-30]. The suppression of Co intensity by annealing suggests the diffusion of Co atoms into the MnGa layer or the diffusion of Mn atoms onto the surface region. Since the 2-nm-thick MgO capping layer is thermally stable at 350°C annealing, the changes of intensity ratio attributes to the atomic diffusion between CoMn and MnGa interface. Therefore, the XAS in Figs. 1 and 3 clearly traces the changes of the compositions in the Co-Mn alloy layer formed by post annealing process as shown in the TEM images in Fig. 2.

Next, the XMCD line shapes are also drastically modulated by annealing contrary to XAS. At the as-grown stage, XMCD line shapes of Co $L$-edge are similar to those of CoMn [18]. The Mn $L$-edge XMCD is composed of both CoMn and MnGa. By the annealing at 250°C, the XMCD intensity of Mn $L$-edge slightly decreases because of the sign reversal in the CoMn layer. The Co $L$-edge XMCD is almost completely disappeared under applying ±1.2 T. In the case of remanent state, the sign of XMCD in Co $L$-edge is reversed. Further annealing at 350°C modulates the Mn XMCD line shapes and the intensity with opposite sign appears dominantly. As shown in Fig. 3(c), the Mn XMCD intensity is suppressed because of antiparallel-coupled Mn sites, and small signals appear like a $Mn^{2+}$ ($3d^5$) XMCD line shape with opposite sign, which might originate from the Mn-O bonding at the interface with MgO or the localized Mn sites in $Co_2MnGa$ alloy because the first-principles calculation suggests the necessary of Coulomb interaction in the Mn sites for the Co-based Heusler alloy [31]. The sign of Co XMCD was already reversed with the metallic line shape. Considering the XAS line shapes, the XMCD in 350°C annealing cannot be explained by the simple sign reversal from the as-grown case, indicating the Mn diffusion and the formation of alternate alloy phase coupled with the MnGa layer antiferromagnetically. The signals from bottom MnGa layer are not detected in the TEY mode because the interface is pushed down by the chemical diffusion. This behavior is different from the case of Co/MnGa interfacial atomic diffusion [12].

Element-specific *M-H* curves at the Mn and Co $L_3$-edge XMCD along the perpendicular direction also show the changes of interfacial magnetic coupling. In the as-grown case, the hysteresis loops with large coercive field and linear slope correspond to the magnetic properties of MnGa and CoMn alloy, respectively. For the Co $L_3$-edge *M-H* curves, the large coercive field is induced through the exchange coupling with the MnGa layer. Since the magnetic fields are applied along out-of-plane direction, a linear slope is originated from the hard axis direction in the in-plane magnetic anisotropy in the CoMn layer [18]. In the case of 250°C annealing, the *M-H* curve in Co $L_3$-edge coupled with the large coercive field is reversed although the slope originated from the CoMn contribution remains unchanged. It suggests that the Co spins facing on the MnGa layer are reversed. Further, the reason for the suppression of Co XMCD signal can be clarified from the *M-H* curve because the summation of hysteresis loop from MnGa and the slope from CoMn compensates the XMCD intensity just in the case of ±1.2 T occasionally. In the case of 350°C annealing, the *M-H* curves for both Mn and Co $L_3$-edge become similar and reversed shapes with large coercive field. It suggests that the uniform Co-Mn alloy layer is produced after the interfacial atomic diffusion. Because of the 30-nm-thick MnGa layer, the large coercive field of Co-Mn alloy still remains as that of MnGa with the strong antiferromagnetic coupling within the XAS probing depth. These suggest that the 350°C annealing promotes the Co-Mn alloy single phase formation and produces the stable reversed PMA compounds from the CoMn/MnGa interfaces.

Considering the above results, we discuss the interfacial reaction at the CoMn alloy/MnGa interface. *First*, the chemical compositions and element-specific spin direction can be discussed. The TEM image clearly shows the formation of Mn and $Co_2MnGa$ layers by 350°C annealing. The XMCD exhibits antiferromagnetic coupling between MnGa and $Co_2MnGa$. The Mn layer formed at the surface region does not contribute to the XMCD signal. Therefore, XAS intensity in Mn L-edge is enhanced as estimated in Fig. 1. The Co XMCD in Fig. 3(c) is similar to the line shape of Heusler alloy $Co_2MnGa$ [32]. From the view point of ternary phase diagram of Co-Mn-Ga [33], it is evident that the Co-rich Heusler alloy exhibits the ferromagnetic phase. *Second*, the interfacial diffusion of Mn atoms has to be considered. Since the bcc CoMn alloy is metastable, the annealing process modulates the phase to fcc structure from the viewpoints of bulk Co-Mn binary phase diagram [34]. Above 150°C annealing, the Mn atom diffusion occurs and the increase of Mn concentration into $Co_{1-x}Mn_x$, resulting in the formation of the fcc phase as reported in the Co-Mn alloys with high Mn concentration [22]. However, in this case, the diffusion of Ga atoms into the CoMn layer prevents the formation of fcc phase, resulting in the formation of bcc Heusler phases in the Co-Mn alloy mixed with the Ga sites, which is evident from the TEM analysis. The formation of Mn layer at the surface region is also the result of high reactivity of Mn atoms. One of the examples of material growth using the reactivity of Mn atoms is the Mn diffusion in the ferromagnetic semiconductor (Ga,Mn)As [35] and the formation of metastable tetrahedral-coordinated structures of MnAs [36]. The Mn doping for metastable phases ca be achieved at the low temperature annealing regions around 200°C without the segregation. *Third*, at the 250°C annealing, the spin flip of Co is the evident for the Mn diffusion from the MnGa layer. However, XMCD shows parallel coupling between Co and Mn in the top layer with opposite direction from MnGa. It suggests the strength of antiparallel exchange coupling $J_{Mn-Mn}$ in MnGa layer is stronger than parallel exchange coupling $J_{Co-Mn}$ in CoMn. According to the DFT calculation, the positive and negative values of $J_{Co-Mn}$ appear depending on the site occupation [25]. The changes of sign in $J_{Co-Mn}$ depending on the ratio of Mn into Co sites are also reported in the CoMnVAl equiatomic Heusler alloy compounds [37]. Therefore, the element-specific magnetization reversal occurs strongly depending on the structural deformation and atomic distance.

Next, we discuss the signs of exchange coupling between Co and Mn sites. In the Slater-Pauling curves, the pair of Co-Mn alloys exhibits the branching behavior because of the appearance of negative $J_{Co-Mn}$, which can explain our experimental results in the case of 250°C annealing. It can be explained by the charge transfer from Co to Mn sites through the orbital hybridization. Furthermore, anti-site defects also weaken the strength of $J_{Co-Mn}$ [25]. Since the $J_{Co-Mn}$ is positive in $Co_2MnGa$, the sign change of exchange coupling in the $Co_2MnGa$/MnGa interface at the 350°C annealing originates from the stronger negative $J_{Mn-Mn}$ in Mn between $Co_2MnGa$ and MnGa than positive $J_{Co-Mn}$ in $Co_2MnGa$ at the interface, resulting in the reversed PMA. These finding accelerates the interfacial materials designing using recent machine learning approaches [38]. *Finally*, the interfacial magnetic behavior in CoMn/MnGa is different from the case in the annealing of Co/MnGa [12]. The reason might originate to the amounts of Mn atoms and the modulation of Mn onto the surface region which trigger the formation of $Co_2MnGa$ at the interface. In order to unveil the interfacial exchange coupling, XMCD with *M-H* curves become a powerful tool and helpful to design the functional materials at the interfaces.

In summary, we studied the magnetic atom diffusion at the interface between 1-nm-thick CoMn and MnGa layer with annealing using TEM and XMCD analyses. We found that the spins in the Mn and Co are coupled parallel to those in the MnGa under the as-grown conditions, while the post-annealing modulates the interfacial magnetic coupling to antiferromagnetic in Co. The element-specific *M-H* curves at each XMCD absorption edge revealed the large coercive fields in Mn and Co through the exchange coupling with MnGa. After the annealing process, the changes of XMCD spectral line shapes indicate that the interfacial reactions promote the formation of the Mn and $Co_2MnGa$ layers, which is deduced from the analysis of multi-probing technique using TEM, XAS, and XMCD.

This work was partially supported by JSPS KAKENHI (Nos. 22H04966 and 21H05000), Yamada Science foundation, and JST-CREST (No. JPMJCR17J5). S. M. thanks to Center for Spintronics Research Network (CSRN). Parts of the synchrotron radiation experiments were performed under the approval of the Photon Factory Program Advisory Committee, KEK (No. 2021G069).

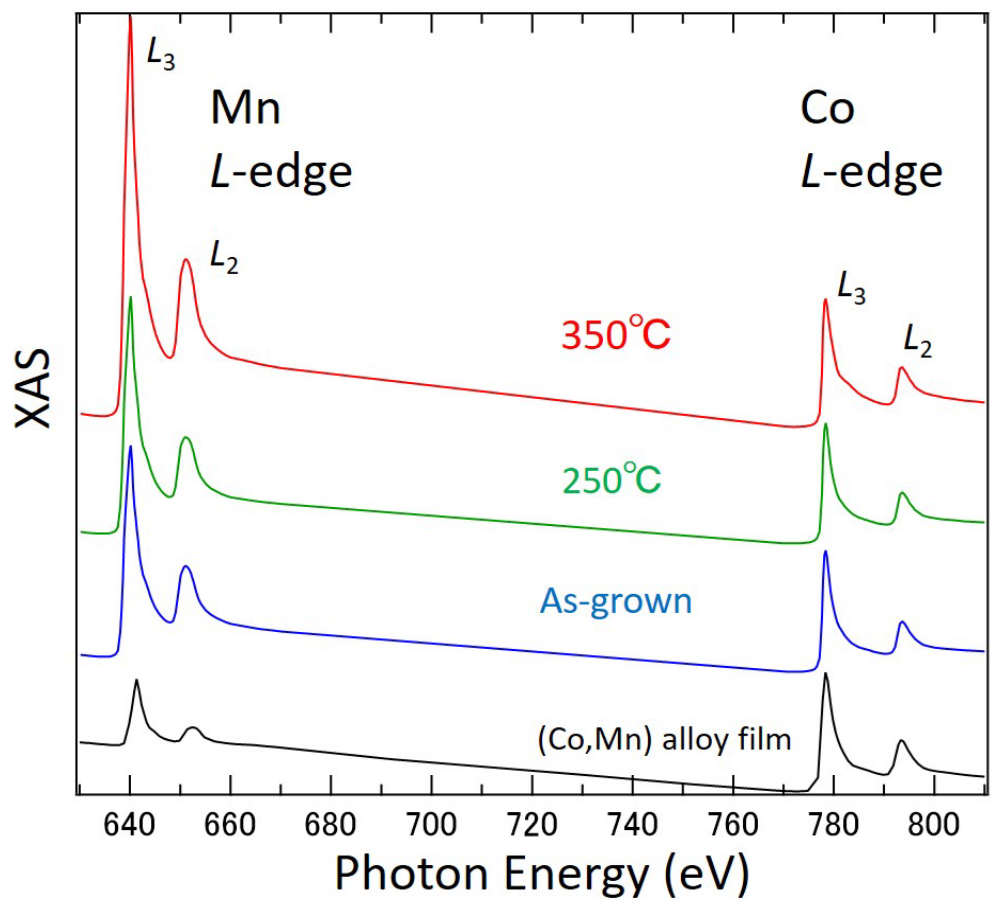

Fig. 1, Wide-range XAS of both Mn and Co $L$-edges in (Co,Mn) alloy/MnGa with annealing temperature dependence: as-grown, 250°C and 350°C annealing cases. As a reference, the same plot of (Co,Mn) alloy without MnGa layer is also displayed. XAS spectra are normalized at the Co $L_3$-edge intensity.

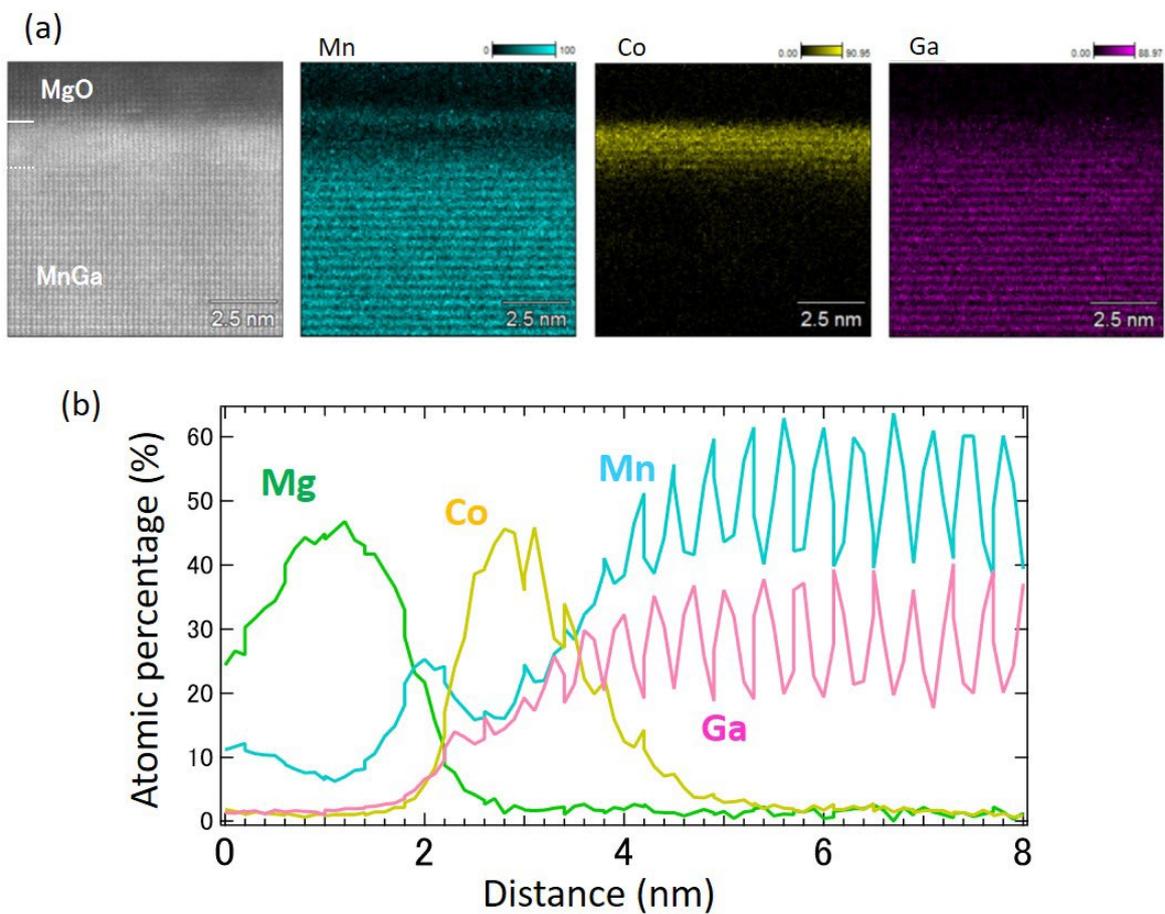

Fig. 2, Cross-sectional profiles of the (Co,Mn) alloy /MnGa after the annealing at 350°C in (a) TEM image and element-specific HAARDF mapping for Mn, Fe, and Ga. (b) Element-specific depth profile by EDX analysis.

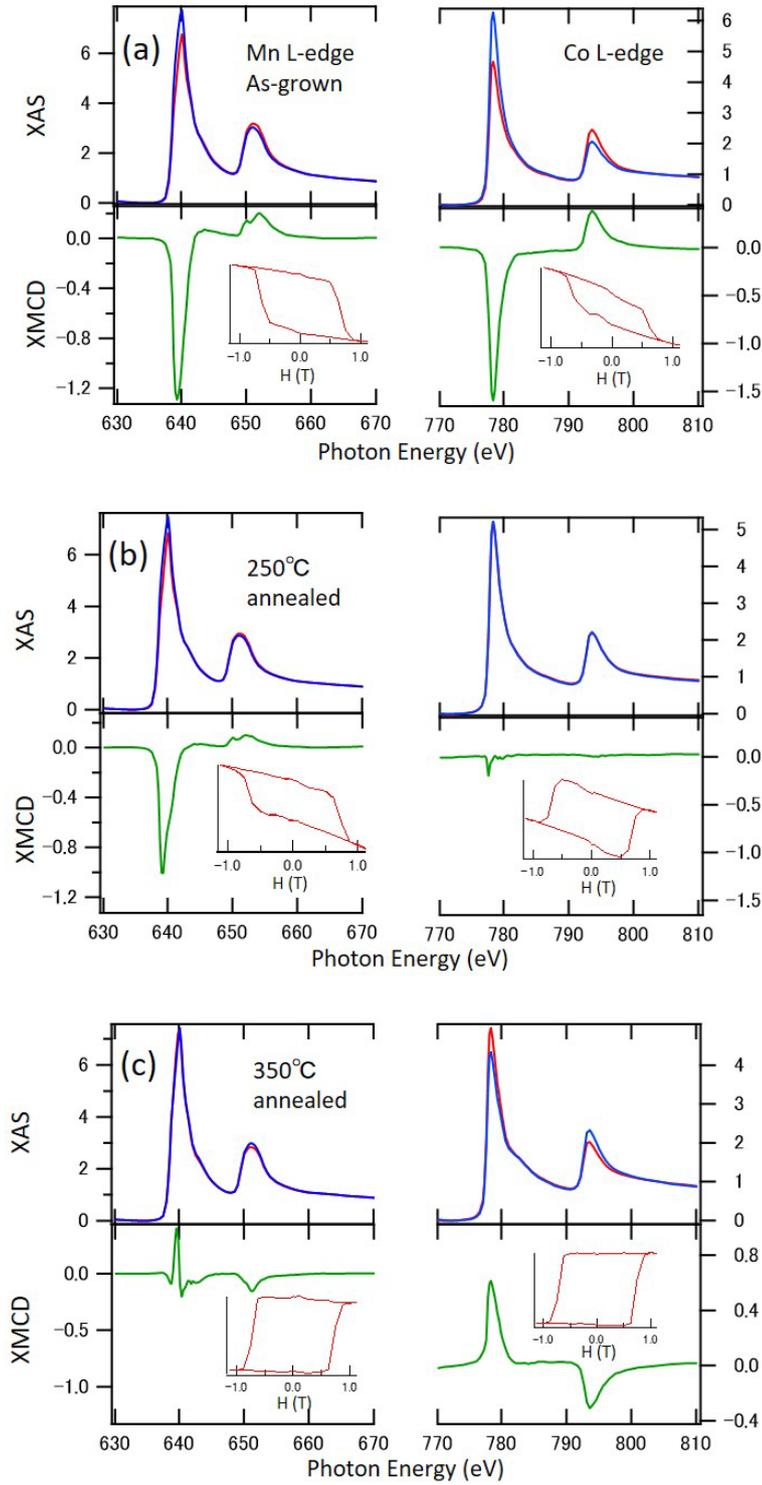

Fig. 3, XAS and XMCD of Mn and Co *L*-edges in (Co,Mn) alloy/MnGa with annealing temperature dependence. (a) As-grown, (b) 250°C annealing, and (c) 350°C annealing cases. Insets show the element-specific hysteresis curves at Mn and Co $L_3$-edges.